\documentclass[journal]{IEEEtran}
\usepackage{cite}

\ifCLASSINFOpdf
  \usepackage[pdftex]{graphicx}
  \DeclareGraphicsExtensions{.pdf,.jpeg,.png}
\else
\fi
\usepackage[cmex10]{amsmath}
\usepackage{widetext}
\usepackage{xspace}

\usepackage{xcolor}

\begin{document}
%
\title{Extended Polarimetric Observations of Chaff using the WSR-88D Weather Radar Network}

%
%
%

\author{James M. Kurdzo,~\IEEEmembership{Senior~Member,~IEEE,}
        Betty J. Bennett,
        John Y. N. Cho,
        and~Michael F. Donovan
\thanks{All authors are with the Massachusetts Institute of Technology Lincoln Laboratory, Lexington, MA 02421 USA email: James.Kurdzo@ll.mit.edu.}
}

%
%


\markboth{IEEE TRANSACTIONS ON RADAR SYSTEMS}%
{Kurdzo \MakeLowercase{\textit{et al.}}: Extended Polarimetric Observations of Chaff using the WSR-88D Weather Radar Network}

%



\maketitle

\begin{abstract}
Military chaff is a metallic, fibrous radar countermeasure that is released by aircraft and rockets for diversion and masking of targets.  It is often released across the United States for training purposes, and, due to its resonant cut lengths, is often observed on the S-band Weather Surveillance Radar -- 1988 Doppler (WSR-88D) network.  Efforts to identify and characterize chaff and other non-meteorological targets algorithmically require a statistical understanding of the targets.  Previous studies of chaff characteristics have provided important information that has proven to be useful for algorithmic development.  However, recent changes to the WSR-88D processing suite have allowed for a vastly extended range of differential reflectivity, a prime topic of previous studies on chaff using weather radar.  Motivated by these changes, a new dataset of 2.8 million range gates of chaff from 267 cases across the United States is analyzed.  With a better spatiotemporal representation of cases compared to previous studies, new analyses of height dependence, as well as changes in statistics by volume coverage pattern are examined, along with an investigation of the new ``full'' range of differential reflectivity.  A discussion of how these findings are being used in WSR-88D algorithm development is presented, specifically with a focus on machine learning and separation of different target types.
\end{abstract}

\begin{IEEEkeywords}
Weather radar, polarimetry, chaff.
\end{IEEEkeywords}

\newcommand{\zh}{$Z_{H}$}
\newcommand{\zv}{$Z_{V}$}
\newcommand{\rv}{$V$}
\newcommand{\rw}{$W$}
\newcommand{\zdr}{$Z_{DR}$}
\newcommand{\rhohv}{$\rho_{HV}$}
\newcommand{\phidp}{$\phi_{DP}$}
\newcommand{\pcdr}{$PCDR$}
\newcommand{\degs}{$^{\circ}$}
\newcommand{\red}[1]{\textcolor{black}{#1}}
\newcommand{\blue}[1]{\textcolor{black}{#1}}

%
\IEEEpeerreviewmaketitle

\section{Introduction}
\begin{figure}[!h]
\centering
\includegraphics[width=3in]{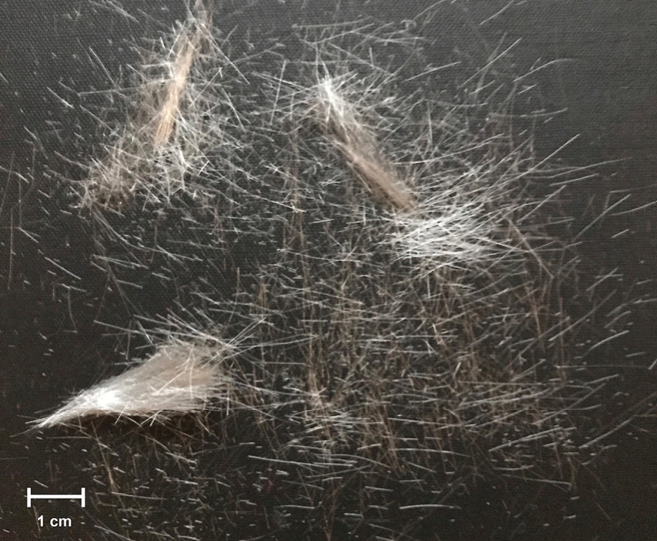}
\caption{Chaff of various cut lengths (from \cite{kurdzo++18}).}
\label{fig_chaff}
\end{figure}

%
%
%
%
\IEEEPARstart{C}{haff}, a military radar countermeasure that consists of metallic fibers cut to specific resonant frequencies, is used to mask aircraft, ships, and missiles from enemy detection \cite{martino12}.  While typically used operationally in the battlefield, chaff releases are common in United States airspace, likely for testing and training purposes (e.g., \cite{murphy++16}).  Given observations of chaff lengths ranging from 1--20 cm on the ground following releases \cite{murphy++16}, these metallic dipoles resonate at frequencies easily detectable by the 10-cm wavelength Weather Surveillance Radar -- 1988 Doppler (WSR-88D) network across the United States \cite{hessemer61,palermo+65}.  An example of chaff fibers and clumps is shown in Fig. \ref{fig_chaff} (from \cite{kurdzo++18}).

\begin{figure*}[!t]
\centering
\includegraphics[width=7in]{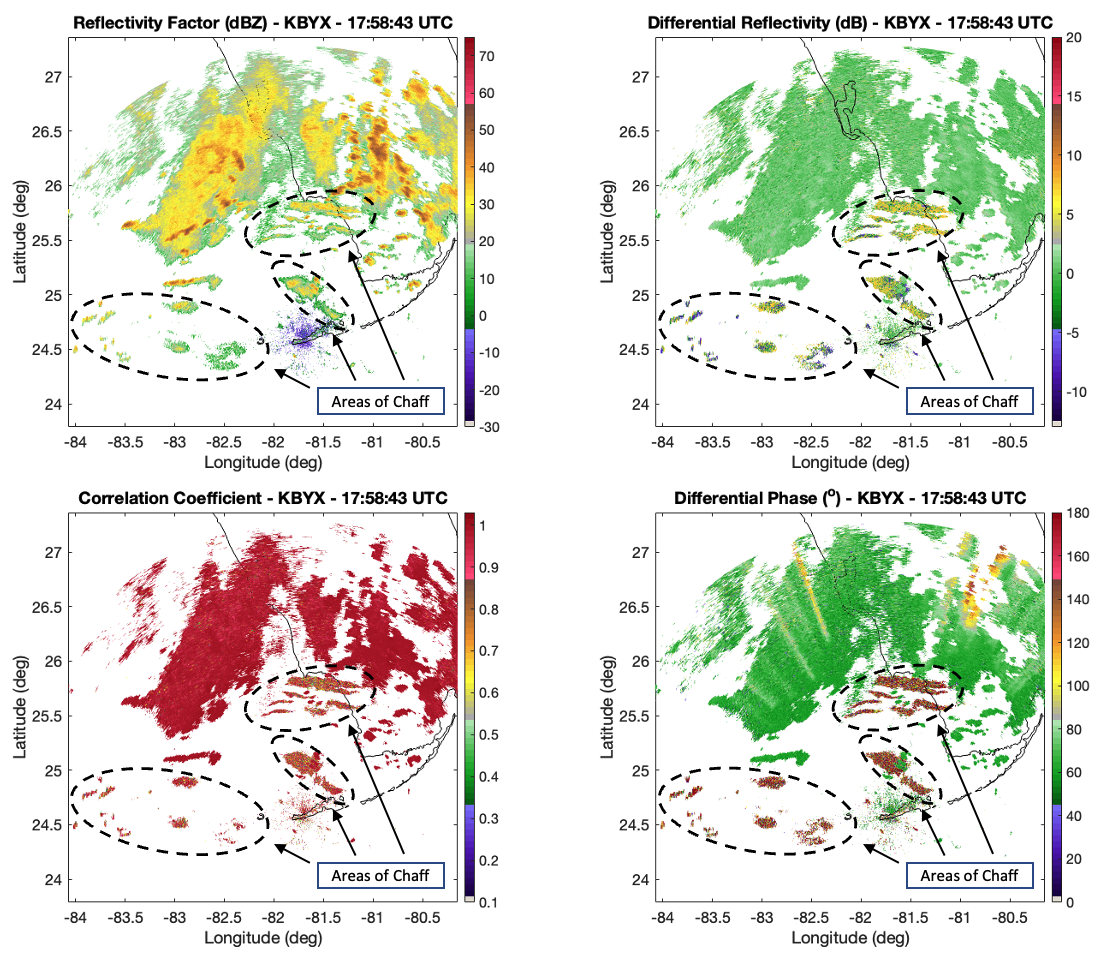}
\caption{An example case of chaff mixed with convection from the KBYX (Key West, FL) WSR-88D on 8 August 2016.  Clockwise from the top left are horizontal reflectivity factor (\zh{}), differential reflectivity (\zdr{}), differential phase (\phidp{}), and co-polar correlation coefficient (\rhohv{})}
\label{fig_example}
\end{figure*}

Chaff is an air motion tracer, meaning that it often flows with the underlying atmospheric velocity field \cite{jessup72,rowland76,sauvageot++82,moninger+87,reinking+96,jung+14}.  While this can theoretically be useful to users of \blue{weather radars}, chaff is generally considered clutter for \blue{these systems} \cite{schuur++03,ryzhkov++05,melnikov++08}.  In many cases, chaff can be nearly co-located with convective storms, and can take on an appearance that makes it exceptionally difficult to differentiate from weather.  This is especially true for radar users that are either inexperienced or do not have access to polarimetric radar variables, which can greatly aid in differentiating between clutter and weather.  As an example, air traffic controllers fall into the latter category, since their displays primarily offer horizontal reflectivity factor (\zh{}) and not polarimetric data \cite{klingle+05}.  An example of chaff mixed with weather is shown in Fig. \ref{fig_example}.  The areas of chaff are nearly impossible to identify by \zh{} alone, but become discernible in the polarimetric fields, especially differential phase (\phidp{}).

As part of \blue{efforts to characterize chaff and sea clutter (another clutter target for weather radar),} several statistical properties of chaff have been examined across hundreds of cases for use in machine learning (ML) classification algorithms.  \blue{These properties include \zh{}, differential reflectivity (\zdr{}), co-polar correlation coefficient (\rhohv{}), and \phidp{}.}  Statistical chaff characteristics were first presented by \cite{zrnic+04}, but only one case in central Oklahoma was examined, leading to a long-held belief that chaff primarily presented only as horizontally oriented fibers, leading to mostly positive \zdr{}.  However, \cite{kurdzo++18} showed more-detailed statistics covering 2.2 million range gates of chaff data, resulting in a deeper understanding of negative \zdr{} in chaff, how often it occurs, and a theoretical basis for why it occurs.

The motivation for this \blue{study} is an upgrade to the WSR-88D that expanded the available \zdr{} range from -7.9 -- 7.9 dB to -13 -- 20 dB.  Since a large focus of \cite{kurdzo++18} was on new observations of \zdr{} in chaff, it is important to characterize \zdr{} with the new, expanded scale.  While exploring these data in a new, approximately 2.8-million range gate dataset collected in 2022, several additional statistical characteristics in chaff that were not covered in \cite{kurdzo++18} were discovered.  \blue{Additionally, through the development of an upgraded/combined sea clutter and chaff detection algorithm, it was found that the currently operational chaff detection algorithm (CDA) \cite{kurdzo++17b} with accuracy of approximately 80\% was able to be improved to nearly 95\% accuracy with the statistics gleaned from this new database when combined with similar data collection of sea clutter cases \cite{kurdzo++23}.}  This paper focuses not only on the newly observed \zdr{} statistics, but also on new observations of other parameters.

This paper is designed to build upon, and augment, \cite{kurdzo++18} for use as a training tool for weather radar users across the weather enterprise.  In addition, ongoing efforts to detect and differentiate chaff and sea clutter \cite{kurdzo++17b} can heavily benefit from these new statistics.  Use of the expanded \zdr{} data, as well as discussions of \zh{} and vertical reflectivity factor (\zv{}) dependence on height, an approximation of circular depolarization ratio (pseudo circular depolarization ratio; \pcdr{}), and different \zh{} and \zdr{} characteristics by WSR-88D volume coverage patterns (VCPs) discussed in this paper will be beneficial to ML algorithms currently under development to discriminate chaff and sea clutter from weather, ground clutter, and biological scatterers.  \blue{Although the paper is focused on data from the WSR-88D network, it is anticipated that the methods and application to ML algorithms will be useful for other radar systems and networks around the world.}

The paper is organized as follows: the data, methods, and description of polarimetric weather radar variables are first discussed.  Next, statistical observations of chaff across the entire dataset are presented.  Finally, a discussion of the findings and conclusions is provided.

\section{Dataset}
The dataset used in this study was collected during an intensive observation period from January -- June 2022 and includes 267 cases (including all tilts from each case) and approximately 2.8 million individual range gates of chaff.  A ``tilt'' is a constant-elevation-angle scan, otherwise known as a plan position indicator (PPI).  A ``case'' is defined as a single volumetric scan at a single point of time from one radar site during a chaff ``event.''  Events are defined as an entire chaff release episode, which can sometimes last over 12 hours.  Cases of chaff sometimes exist from separate radar sites at a similar time to another radar site(s), but, in most cases, only one time per event per radar is included in the database.  A small number of events (fewer than 10) contain two cases that are at least 8 hours apart due to significant changes in chaff structure and statistics over a long-duration release.  This means that, for the majority of events, the dataset does not contain multiple volumes from the same radar in the same event, making the database spatiotemporally diverse, especially compared to \cite{kurdzo++18}.  In \cite{kurdzo++18}, some cases were from similar times within the same event and from the same radar.  Additionally, in the present dataset, there is a significant increase in the diversity across radar sites spatially.  Cases from 71 sites across the CONUS are included, with the most from one radar being 32 from KNKX (San Diego, CA).

It should be noted that despite the much larger number of cases in this paper compared to \cite{kurdzo++18} (267 versus 75), only a modest increase in the number of range gates used for analysis is realized (2.8 million versus 2.2 million).  This is due to the fact that a wider range of sites and ``types'' of cases were included in the dataset.  In \cite{kurdzo++18}, many of the cases were relatively large spatially since they were straightforward to identify.  \blue{The currently operational CDA} was developed with the statistics from \cite{kurdzo++18} and operationalized \red{on the} WSR-88D in 2021 \blue{\cite{kurdzo++17b}}.  This algorithm was used \blue{only} to \blue{\textit{identify}} cases across the entire WSR-88D fleet, including smaller spatiotemporal cases with varying statistics.  \blue{It is important to note that while the existing CDA was used to identify \textit{possible} cases, all data in this study were collected manually via the description below.  No automation of case selection was used.}

The spatial distribution of sites and the temporal distribution across six months in the present dataset makes it an arguably better representation of chaff characteristics across the CONUS than that in \cite{kurdzo++18}, while also including the added \zdr{} range, discussion of \zv{}, new \pcdr{} calculations, and separation by height and VCP.  All statistics are calculated using the Open Radar Product Generator Simulator (ORPGSim), a MATLAB clone of the operational WSR-88D ORPG.  ORPGSim provides nearly identical results to the operational ORPG for generating products, but offers the flexibility of simplified processing and workflows for collection and analysis of large amounts of data.  As part of this software package, a region-of-interest tool was used to manually truth range gates of chaff using a subject matter expert-defined polygon drawn around clusters of chaff at each tilt within a volume.  A description of this collection technique is detailed in Section 2c of \cite{kurdzo++18}.  Every tilt of every case was manually truthed by a human, and the associated masks were used to isolate chaff range gates.  \red{It should be noted that there is no \textit{actual} ground truth of chaff in any of our cases (i.e., no definitive proof that what we are seeing is indeed chaff).  The only known case with literal ground truth was presented in \cite{murphy++16}, where chaff was found on the ground after a very concentrated chaff release.  Our approach is based on the fact that no other known target type displays these signatures, and the signatures that serve as our ``truth'' are relatively well-accepted in the broader literature.}

In order to further refine the chosen range gates, all \rhohv{} values $\geq$0.97 were removed from the final dataset.  Additionally, all range gates with a standard deviation of \phidp{} less than 10\degs{} were removed. The standard deviation of \phidp{} in this context is an ORPG-defined product that uses unwrapped \phidp{} values.  This is a different approach from \cite{kurdzo++18}, and is meant to remove range gates of chaff that are mixed with, and dominated by, weather.  \blue{The values chosen for censoring are based on the fuzzy logic membership functions in the Inanimate class of the Hydrometeor Classification Algorithm (HCA; \cite{park++09}) discussed in \cite{kurdzo++20}}.  \red{These values were originally determined using a genetic algorithm to optimize the membership functions using a series of chaff, sea clutter, and ground clutter cases.  They were found to be the optimal combination of parameters to identify these target types and avoid categorizing weather and other clutter as chaff or sea clutter.}  These more-stringent restrictions also account for a non-trivial decrease in the number of range gates in the database used for analysis.

Similar to \cite{kurdzo++18}, an unfolded ``full'' \phidp{} product was mimicked in ORPGSim, allowing \phidp{} values up to 360\degs{}.  Values 60\degs{} and below were not used due to the system initial phase generally being set to 60\degs{}, resulting in a large spike where \phidp{} values at or below 60\degs{} are cut off.  This invariably misses some folded \phidp{} values, but the number of range gates that fit into this category is exceptionally low.  For height calculations, a standard 4/3 Earth's radius model is used to calculate the beam height at each range gate based on range and elevation angle \cite{doviak+93}.

In addition to the aforementioned radar variables (\zh{}, \zdr{}, \rhohv{}, \phidp{}), also presented in \cite{kurdzo++18}, this study also discusses \zv{}, Doppler spectrum width (\rw{}), and \pcdr{}.  The formulation used for \pcdr{} is an approximation of the circular depolarization ratio ($CDR$) for simultaneous transmission/reception of horizontal/vertical polarization (SHV) radars defined in \cite{ryzhkov++17}:

\begin{equation}
PCDR =\frac{1+Z_{DR}^{-1}-2\rho_{HV}Z_{DR}^{-1/2}}{1+Z_{DR}^{-1}+2\rho_{HV}Z_{DR}^{-1/2}}
\label{eq.pcdr}
\end{equation}

\blue{It is important to note that in this study, the inputs of \zdr{} and \rhohv{} to (\ref{eq.pcdr}) are in \red{derived product} form, as with the rest of the analyzed data.  Within the WSR-88D processing scheme, \red{derived products} consist of post-processed ``products'' that are outputs from the ORPG; \red{derived products} are the input source to most downstream ORPG algorithms such as the HCA.  The primary \red{derived} products are modified versions of the \red{base radar variable} estimates, including processing such as recombination from 0.5\degs to 1.0\degs and other data quality assurance measures.}

$CDR$ has been shown to be useful for discriminating between meteorological and non-meteorological targets \blue{\cite{ryzhkov++02b,ryzhkov++17,kilambi++18,michelson++20,perez++23}} and, most relevantly to this study, between hydrometeors and chaff \cite{moninger+87}.  Given that it is not a ``native'' variable in SHV radars, it is rarely used operationally \cite{ryzhkov++17} and is an unused tool in algorithmic development \red{on the WSR-88D}.  Although \cite{jameson87} showed that $CDR$ can be expressed in terms of linearly polarized radar parameters, \cite{moninger+87} noted that this formulation is quite sensitive to propagation effects, and instead proposed \pcdr{} (or what they called depolarization ratio), which is more robust against propagation effects.  With work underway to differentiate chaff from other target types, a discussion of \pcdr{} and its relation to other variables of interest is considered warranted in this study.

\section{Statistical Observations}
This section focuses on the broad statistical properties of chaff within the newly collected dataset from 2022.  A deeper discussion on selected topics is presented in the following section.  For each figure shown, all 2.8 million range gates across all 267 cases are included in the histograms.  Direct comparisons are made with the dataset in \cite{kurdzo++18} when appropriate.

\begin{figure*}[!t]
\centering
\includegraphics[width=7in]{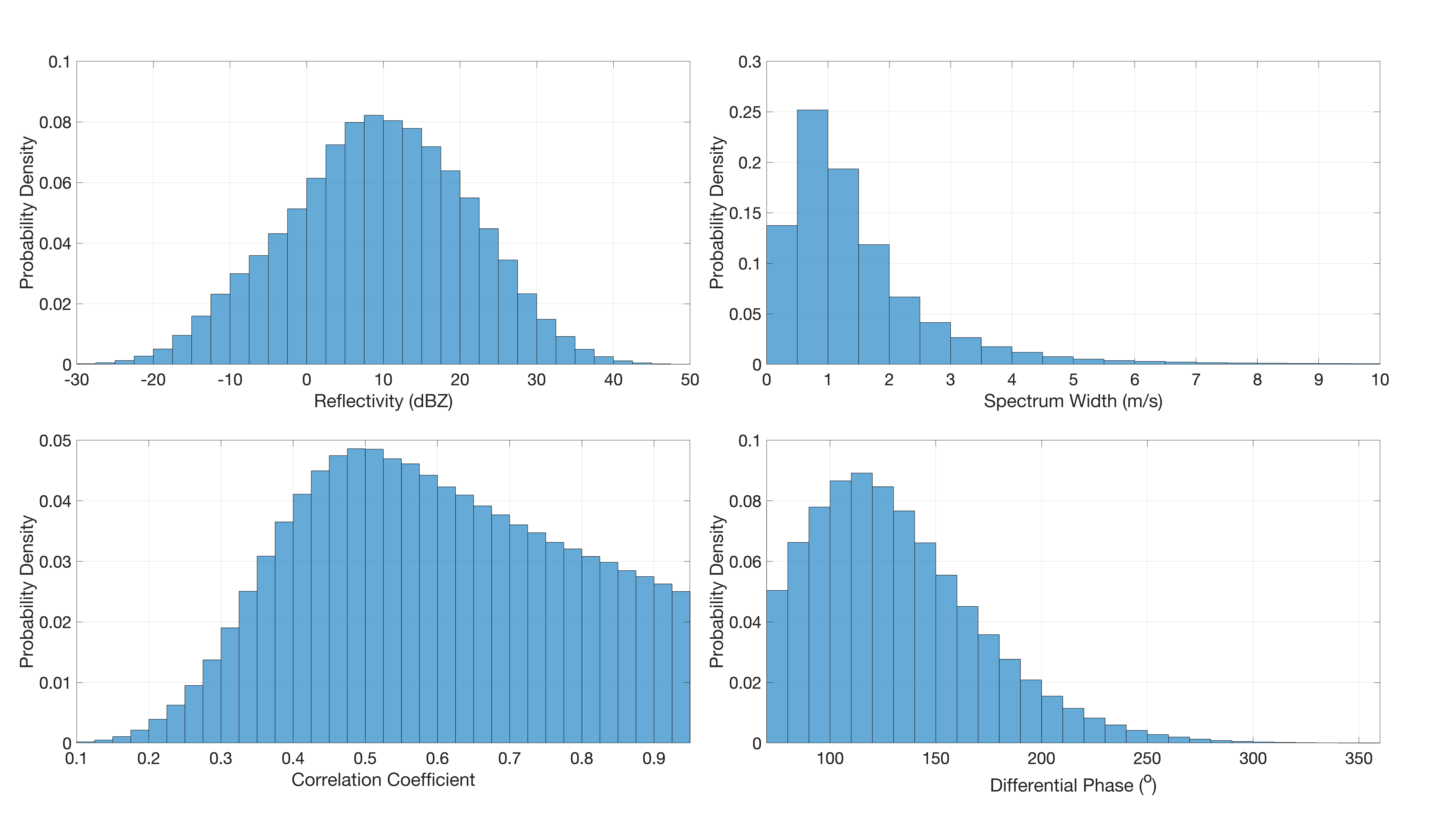}
\caption{Histograms of approximately 2.8 million range gates of chaff from 267 cases.  From top left, counterclockwise: horizontal reflectivity factor (\zh{}), spectrum width (\rw{}), differential phase (\phidp{}), and co-polar correlation coefficient (\rhohv{}).}
\label{fig_summary}
\end{figure*}

Histograms of \zh{}, \rw{}, \rhohv{}, and \phidp{} are shown in Fig. \ref{fig_summary}.  \zh{} shows a \blue{roughly Gaussian} shape centered just below 10 dBZ.  The observed distribution differs from the ``plateau'' seen in Fig. 3 of \cite{kurdzo++18}, but is centered in a similar range.  The higher concentration of values from -10 dBZ -- 10 dBZ is thought to be due to the addition of more cases with varying spatial extents.  As opposed to the dataset in \cite{kurdzo++18}, which chose cases that were often at their peak spatial extent (i.e., the most easily identifiable cases), the present dataset includes a much-higher number of cases that have relatively small spatial footprints, both vertically and horizontally.  Anecdotally, through extensive data collection and observation, cases with smaller chaff cloud sizes tend to have lower reflectivity, which lends credence to the observed histogram.

The histogram of \rw{} was not included in \cite{kurdzo++18}, but is shown here due to its importance in the work from \cite{kurdzo++20}, which utilized statistics from \cite{kurdzo++18} to develop a new hydrometeor class (called ``Inanimate'') for the WSR-88D HCA.  It was found in \cite{kurdzo++20} that exceptionally low values of \rw{} were present in all of the target types encompassed by the Inanimate class, which include chaff, sea clutter, combustion debris, and radio frequency interference.  However, in recent attempts to separate chaff from sea clutter, there are distinct differences in the peak and tail of \rw{}, making these findings relevant at the current time.  It makes intuitive sense that chaff would have a low \rw{} distribution since most cases collected did not overlap with convective storm cells where high \rw{} values can be expected.  Increased values of \rw{} do seem to be present in cases of chaff near or adjacent to convection (not shown), but the tail drops off quickly after 3 -- 4 m/s.  \blue{It should be noted that \textit{most} cases in the database consist of chaff-only scatterers, so the true values in the \rw{} histogram may extend slightly higher in cases where chaff is mixed with convective weather.  However, in most cases, the dominant scatterer (be it chaff or weather) is what is ``detected'' by the radar.  Of course, if chaff is detectable but mixed with weather, we would expect to see elevated \rw{} estimates.}

In \rhohv{} and \phidp{}, the distributions are largely similar to those shown in \cite{kurdzo++18}.  The differences are that \rhohv{} values greater than or equal to 0.97 were discarded in this dataset due to the proximity of weather in some cases, and values of \phidp{} less than or equal to 60\degs{} were discarded due to the setting of system initial phase.  Range gates where the standard deviation of \phidp{} was less than or equal to 10\degs{} were also discarded, which slightly alters the \phidp{} distribution.  \blue{The standard deviation of \phidp{} was calculated in the identical fashion to the ORPG, which creates a ``hidden'' product that is part of the operational HCA.  This product is what is referred to as a texture field in \cite{park++09}, which is calculated using a comparison between a sliding smoothing window and the original data.  The exact filter values are detailed in \cite{park++09}.}

For \rhohv{}, a peak near 0.5 with a relatively sharp dropoff toward 0.1 is seen.  A much more muted drop in distribution of \rhohv{} is seen towards the highest values at 0.96.  It should be noted that, as in \cite{kurdzo++18}, there are cases of chaff range gates that extend to the maximum \rhohv{} \red{estimate} of \red{1.03}, but they were discarded for this dataset.  \red{Note that in the WSR-88D processing of derived products, some \rhohv{} estimates can be as high as 1.03.  These values were clipped and removed in the histograms.} For \phidp{}, as with the dataset in \cite{kurdzo++18}, a peak around \blue{100 -- 120\degs{}} is seen, with a gradual dropoff toward 200\degs{} and a long tail to 360\degs{}.  Unlike Fig. 3 in \cite{kurdzo++18}, folded values that end up below 60\degs{} are not included.

\blue{Typically, \phidp{} is not a tool directly used in weather radar analysis; rather, its path derivative, specific differential phase ($K_{DP}$), is used.  However, in many dual-polarimetric weather radars, the WSR-88D included, $K_{DP}$ is not calculated for \rhohv{} values under a given value (0.9} \red{for the WSR-88D}\blue{) due to the unlikelihood that rain is present.  Therefore, on many radars, $K_{DP}$ is not calculated in instances of chaff.  It turns out that \phidp{} is itself very useful for identifying chaff due to its highly stochastic nature compared with the relatively smooth gradients seen in weather (as seen in Fig. \ref{fig_example}).  This appearance is due to the fact that chaff, if treated as a dipole, has} \red{zero differential backscatter phase. The combination of receiver differential phase as the primary contributor and high measurement error due to very low cross-correlation coefficients \cite{zrnic+04} makes for this hallmark stochastic presentation in phiDP \cite{kurdzo++18}.  In cases when the standard deviation of \phidp{} is less than 10\degs{}, we have seen, both qualitatively and quantitatively, that chaff is exceptionally unlikely to be present, \textit{or}, weather is the dominant scatterer in the resolution volume.}

\begin{figure*}[!t]
\centering
\includegraphics[width=7in]{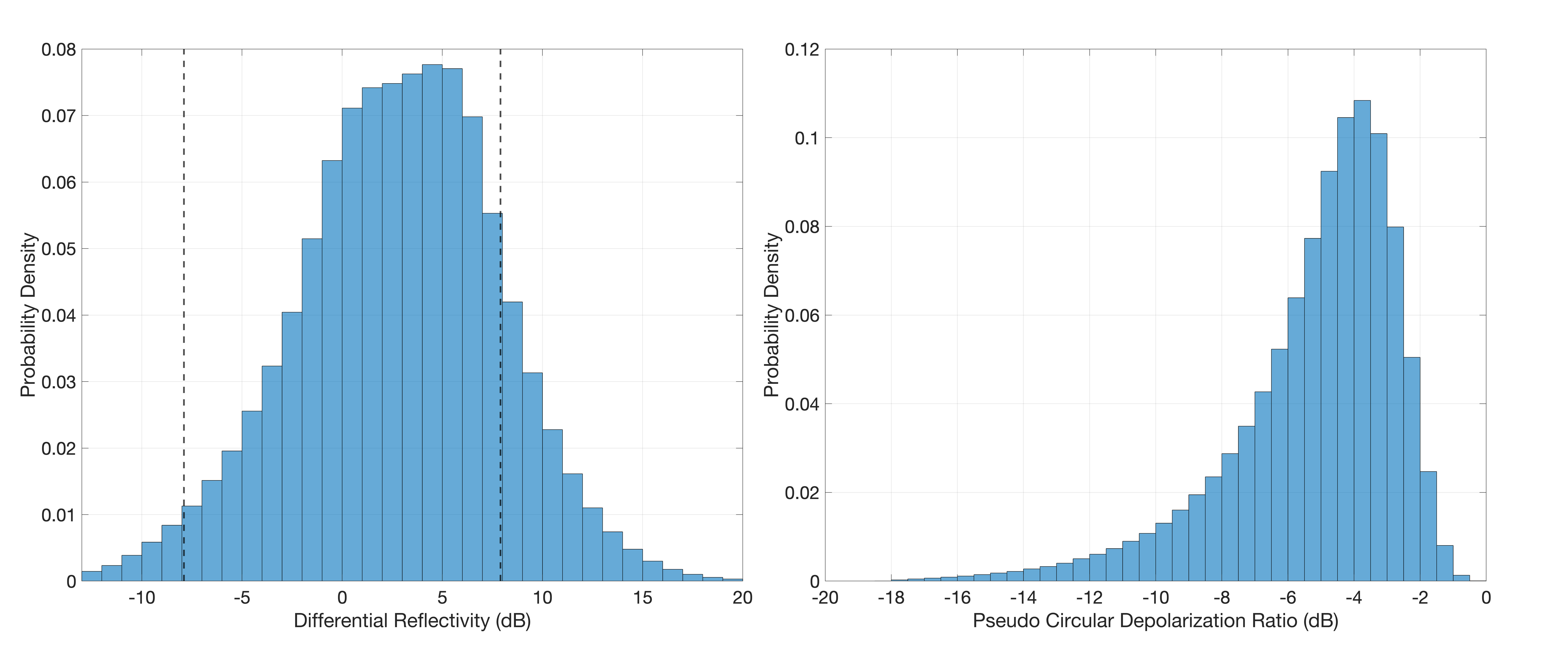}
\caption{Histograms of approximately 2.8 million range gates of chaff from 267 cases. Left: differential reflectivity (\zdr{}); right: pseudo circular depolarization ratio (\pcdr{}) from \cite{ryzhkov++17}.  Vertical dashed black lines indicate the previous \zdr{} range of -7.9 -- 7.9 dB.}
\label{fig_zdr_dr}
\end{figure*}

With the latest \blue{software processing upgrades to the WSR-88D (as of late 2021)}, every case in the present 2022 dataset is able to provide \zdr{} information between -13 -- 20 dB.  Previously, values that fell beyond \mbox{-7.9} -- 7.9 dB were clipped.  In \cite{kurdzo++18}, those clipped values were eliminated.  In Fig. \ref{fig_zdr_dr}, the full range of \zdr{} is shown with the dashed vertical black lines indicating the previous -7.9 -- 7.9 dB \zdr{} range.  Within the previous range, the distribution is a bit different than that seen in \cite{kurdzo++18}.  A more plateaued shape is seen in the 2022 dataset, with relatively flat probability densities between 0 and 5 dB.  In \cite{kurdzo++18}, there was a peak around 1.5 dB and a slow dropoff toward the 7.9 dB limit, \blue{while here, the peak is at 4 dB.  The relatively plateaued shape is in stark contrast to the sharper peak at 1.5 dB in \cite{kurdzo++18}, demonstrating a consistent distribution of \zdr{} values between roughly 0 -- 5 dB.  This observation of a shift toward higher \zdr{} values is thought to be indicative of a more-diverse dataset in the present study.  For example, with a wider spatial range of data (i.e., more sites across the country), different types of chaff may be used, leading to a possible shift in distribution.  However, it is not completely clear why this shift is seen in the 2022 data.}

In Fig. \ref{fig_zdr_dr}, the dropoff after 5 dB is sharp and rapid, decreasing by nearly a factor of two before reaching 8 dB.  However, most importantly, a non-trivial percentage of the distribution (17\%) is observed outside of the original bounds.  This is particularly true on the positive side ranging from 7.9 -- 20 dB.  A total of nearly 15\% of all chaff range gates fall above 7.9 dB.  Additionally, a small tail also extends from -13 -- -7.9 dB, showing additional information gleaned for strongly vertically oriented fibers with the new \zdr{} range.

\blue{It should be noted that \zdr{} calibration is a known issue with the WSR-88D system \cite{zrnic++06}, leading to biases that can sometimes exceed $\pm$0.5 dB \cite{richardson++17}.  While these biases can be monitored using techniques such as Bragg scatter detection, they are not recorded as a definitive value in WSR-88D data (and can change over even short periods of time) \cite{richardson++17}.  Since correcting for \zdr{} biases is impractical for this study, it is assumed that the biases are small enough to still get the general trend of \zdr{} values.  It is also important to consider that the distribution of \zdr{} values in this study represent the data used to train the updated CDA in \cite{kurdzo++23}.  This generally favors algorithm performance, since calibration is not accounted for in real-time data streaming either (i.e., as data are fed to the algorithm for processing and display).}

The distribution of \pcdr{}, calculated using the approximation from \cite{ryzhkov++17} shown in (\ref{eq.pcdr}), has not been explored before in a large number of chaff cases.  Several studies have shown various types of clutter to fall within the range of -10 -- -1 dB, while weather is generally much more negative \cite{ryzhkov++17}, since spherical particles reflect much more strongly in the opposite circular polarization sense than in the same sense.  In Fig. \ref{fig_zdr_dr}, the \pcdr{} histogram fits this expectation, with a peak close to -4 dB, a sharp dropoff to -1 dB, and a long tail out to approximately -18 dB.  Further discussion on these findings and their importance to algorithmic development is provided in the next section.

\begin{figure}[!t]
\centering
\includegraphics[width=3.5in]{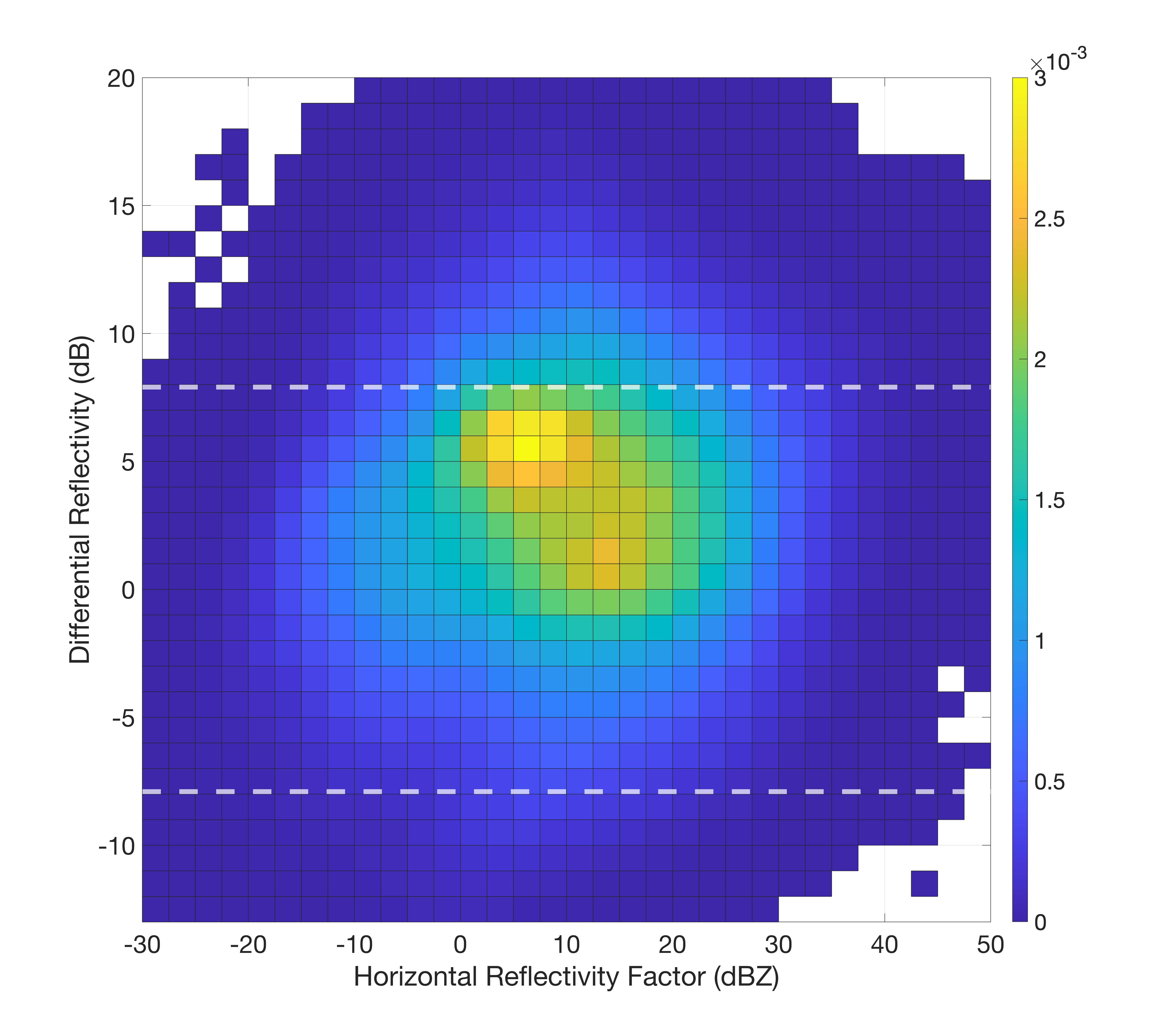}
\caption{Distribution of chaff range gates between horizontal reflectivity factor (\zh{}) and differential reflectivity (\zdr{}).  Horizontal dashed white lines indicate the previous \zdr{} range of -7.9 -- 7.9 dB.}
\label{fig_z_zdr}
\end{figure}

Given the advent of a new range for \zdr{}, it was prudent to compare \zh{} versus \zdr{} in a similar fashion to Fig. 5 from \cite{kurdzo++18}.  The results are shown in Fig. \ref{fig_z_zdr}, with \zh{} along the abscissa and \zdr{} along the ordinate.  Similar to the differences in \zh{} distribution between Fig. \ref{fig_summary} and \cite{kurdzo++18}, the distribution in Fig. \ref{fig_z_zdr} is skewed toward lower \zh{} values compared to \cite{kurdzo++18}.  Additionally, the plateaued nature of the \zdr{} distribution in Fig. \ref{fig_zdr_dr} is evident in Fig. \ref{fig_z_zdr}.  When combined, there is a range of distinct hot spots from 5 dBZ / 5 dB (for \zh{} and \zdr{} respectively) to 15 dBZ / 0 dB (also for \zh{} and \zdr{} respectively).  While this is not surprising given the histograms in Figs. \ref{fig_summary} and \ref{fig_zdr_dr}, two additional interesting features are apparent.  First, at \zdr{} values beyond 7.9 dB, the \zh{} values tend to sit between 0 and 20 dB.  This indicates that the extension of \zdr{} to higher values correlates with a wide range of \zh{} values.

\begin{figure*}[!t]
\centering
\includegraphics[width=7in]{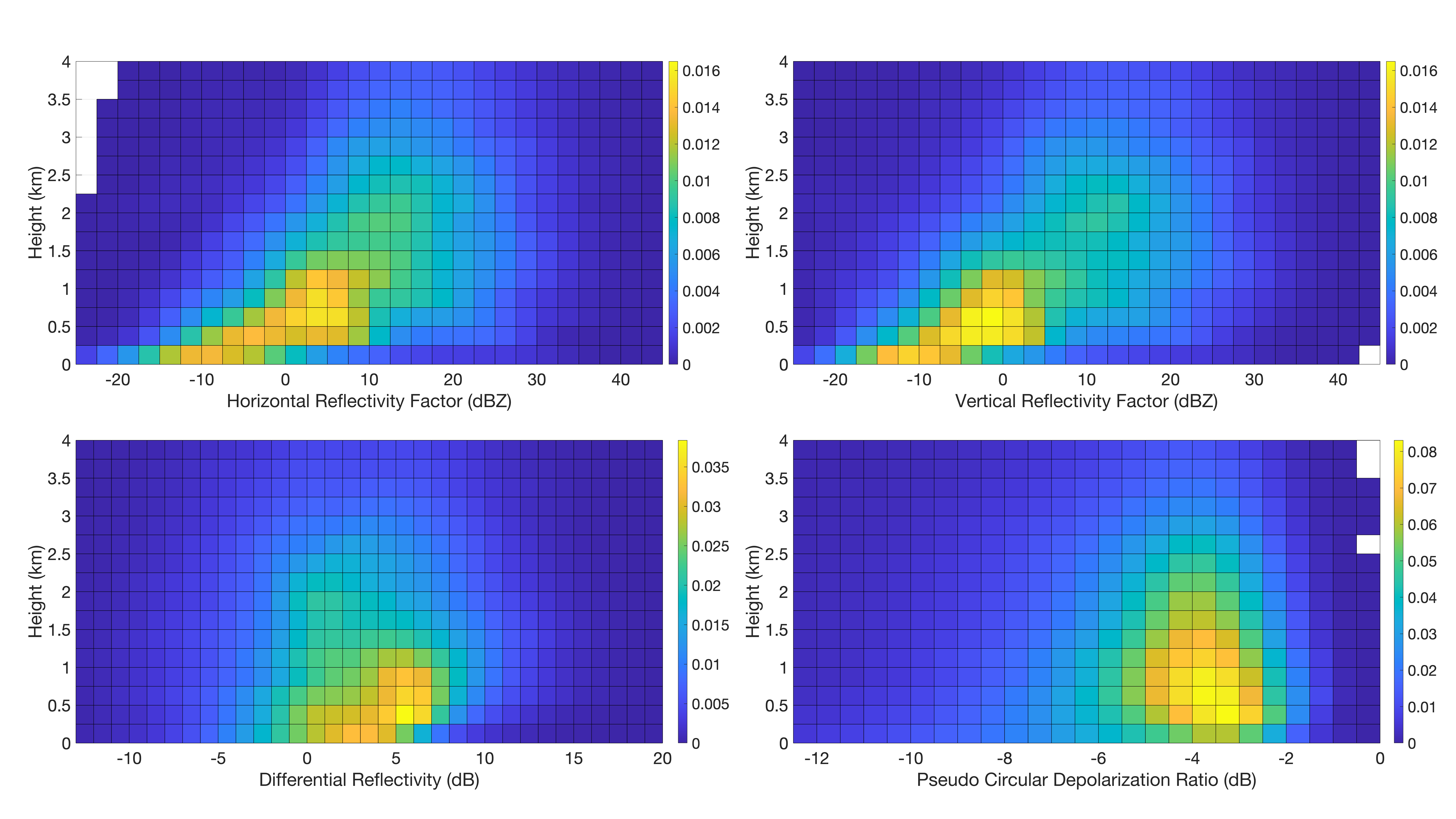}
\caption{Height-based chaff distribution of horizontal and vertical reflectivity factors (\zh{} and \zv{}; top row, from left to right), as well as differential reflectivity and pseudo circular depolarization ratio (\zdr{} and \pcdr{}; bottom row, from left to right).  Heights are relative to the radar's elevation.}
\label{fig_z_zdr_dr_height}
\end{figure*}

In \cite{kurdzo++18}, the only examples of height dependence were presented for \zdr{} and were focused on individual case studies.  Violin plots were shown for several ranges of heights spanning eight different case studies.  \zdr{} was also explored temporally throughout the lifetime of a chaff cloud's fallout for the same eight cases.  However, no holistic view of \zdr{} with height was explored, and no statistics of \zh{} were discussed in a spatiotemporal sense.  In Fig. \ref{fig_z_zdr_dr_height}, the dependence of \zh{}, \zv{}, \zdr{}, and \pcdr{} with height are shown.  Both \zh{} and \zv{} show trends of increasing with height, although \zv{} peaks at slightly lower values at a similar height as \zh{} (roughly 750 m above radar level).  A discussion on this effect and what it implies for algorithm development is provided in the following section.

Also in Fig. \ref{fig_z_zdr_dr_height}, the dependence of \zdr{} on height is shown.  In general, the \zdr{} distribution above 250 m stays relatively similar with height, mostly matching the histogram in Fig. \ref{fig_zdr_dr} and agreeing with what would be expected from the \zh{} and \zv{} distributions in the top row of Fig. \ref{fig_z_zdr_dr_height}.  However, a negative shift in distribution is apparent at lower heights, which would agree well with the case studies in \cite{kurdzo++18} where \zdr{} values close to zero were seen at the lowest levels and \zdr{} values between 3 -- 6 dB were more common at higher elevations.  Finally, \pcdr{} displays a relatively consistent distribution with height, with a strong focus around -3 to -4 dB.  There appears to be a slight decrease in \pcdr{} with height, near \mbox{-0.5} dB per km between 250 m and 2 km.  Additional discussion on these observations follows in the next section.

\begin{figure}[!t]
\centering
\includegraphics[width=3.5in]{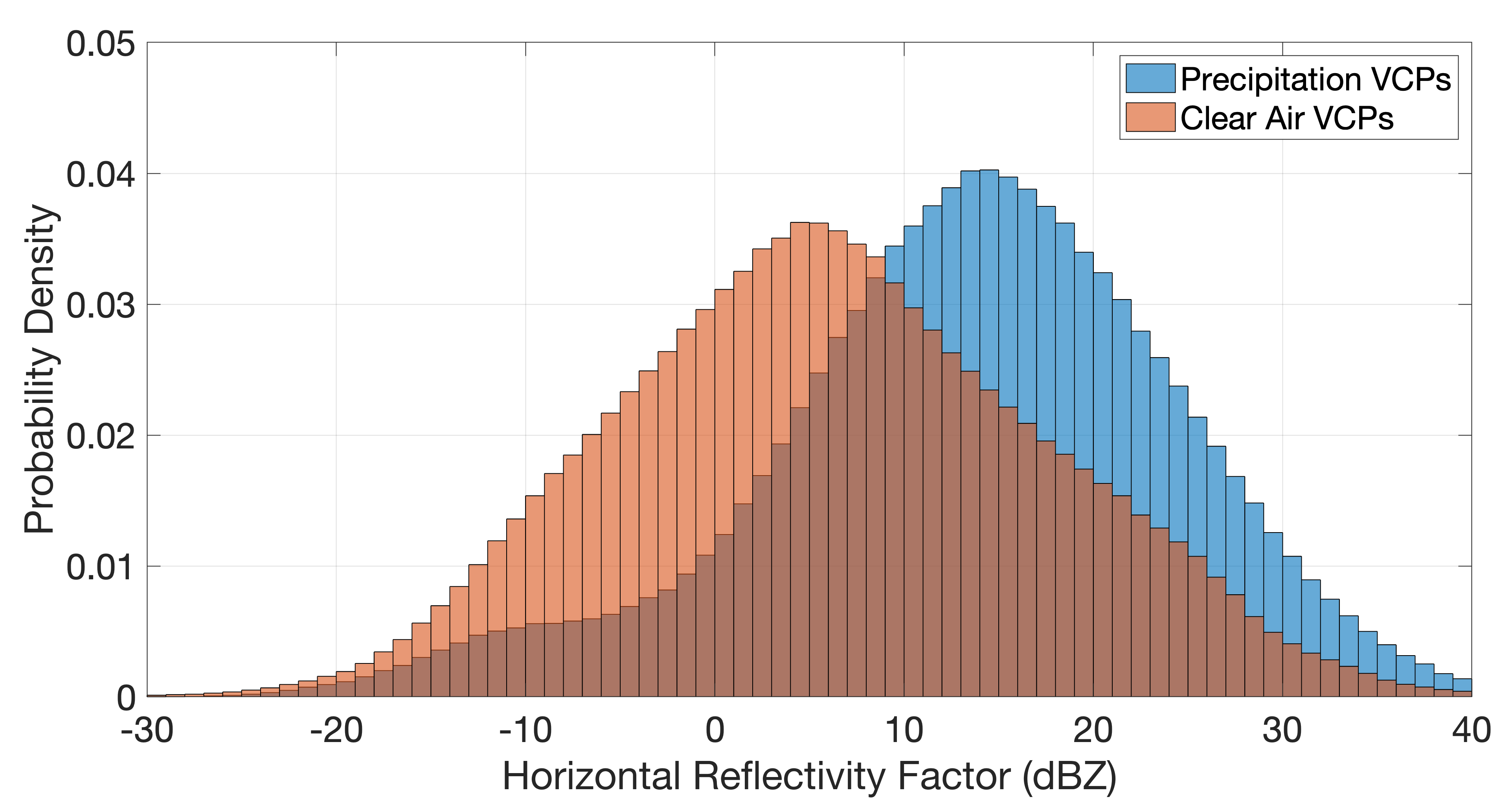}
\caption{\red{Histograms of chaff horizontal reflectivity factor (\zh{}) broken out by precipitation and clear air volume coverage patterns (VCPs).}}
\label{fig_z_zdr_vcp}
\end{figure}

Finally, \red{differences were noted in the database when comparing statistics between different VCPs.}  \blue{VCPs are defined by the scan pattern of the radar, including (but not limited to) the number/exact value of tilts, pulse repetition frequency, and scan rate.  In general, precipitation VCPs have more tilts but scan at higher rates, resulting in both faster volumetric and single-elevation update rates.  Despite their name, clear air VCPs are often used in cases of light stratiform precipitation, especially snow.  VCP 31, a clear air VCP, is a unique scanning mode for the WSR-88D in that it is the only VCP that uses a longer pulse (4.7 $\mu $s) than the other VCPs (1.57 $\mu $s).}

In order to investigate \red{these observations}, each range gate within each case was separated into the current VCP at the time of the case.  Histograms of \zh{} for two sets of VCP combinations are detailed in Fig. \ref{fig_z_zdr_vcp}: \blue{precipitation versus clear air VCPs}.  In Fig. \ref{fig_z_zdr_vcp}, the clear air VCPs show a much lower peak \zh{} value in their distribution relative to the precipitation VCPs.  The precipitation VCPs are centered nearly 10 dBZ higher than the clear air VCPs, but have a long, low tail out to \nobreakdash-30 dBZ (compared with a more-gradual falloff at low \zh{} values in the clear air VCPs).  Thoughts behind these observations are offered in the following section.

\section{Discussion}
A more-detailed discussion of three selected statistics is provided in this section, including the distribution of \pcdr{}, observations of \zh{}, \zv{}, \zdr{}, and \pcdr{} by height, and the distribution of \zh{} and \zdr{} broken out by VCP.

\subsection{Depolarization Ratio Distribution}

As chaff is considered an air motion tracer, and is most-often found in laminar airflow (i.e., not near convection), chaff is oriented horizontally \textit{most} of the time (as evidenced in Fig. \ref{fig_zdr_dr}).  It is known from \cite{kurdzo++18} that there is a period of time during chaff fallout, usually during the beginning, that features vertically oriented fibers, contributing to most of the negative \zdr{} values in the histograms in both \cite{kurdzo++18} and the present study.  Additionally, as discussed in \cite{zrnic+04}, the linear depolarization ratio ($LDR$) in chaff is expected to be relatively high compared to precipitation.

\cite{zrnic+04} assumed that the angle between the horizontal plane and the chaff axis was distributed uniformly between zero and a maximum angle called the flutter angle.  With this approach, a flutter angle of zero corresponds to all chaff being horizontally oriented (\zdr{} = +$\infty$), while a flutter angle of 90\degs{} corresponds to a uniform distribution of chaff orientation (\zdr{} = 0 dB).  Their Hertzian dipole model using $LDR$ (their Fig. 3) showed that $LDR$ values are asymptotic approaching just above -5 dB as the flutter angle approaches 90\degs{}.  In fact, the model only dips to about -10 dB at a flutter angle of 30\degs{}, which corresponds to an approximate \zdr{} of 15 dB.  This suggests that, in the presence of chaff, it would be expected to see values of $LDR$ between -10 and -5 dB, but more extremely horizontally oriented chaff could dip to lower $LDR$ values (\nobreakdash-13 dB for a \zdr{} value of 20 dB).  However, $LDR$ cannot be measured by a radar operating in SHV mode, like the WSR-88D. Thus, we consider the use of $CDR$ instead, which can be approximated using SHV-derived variables via the \pcdr{} formulation.

For a radar operating in circular polarization mode, the $CDR$ is the received cross-polarized power divided by the co-polarized power. $CDR$ is expected to be an excellent discriminator between hydrometeors and chaff because a perfect sphere reflects all signal in cross-polarized form ($CDR$ of +$\infty$), whereas a resonant dipole (an approximation of a chaff strand) has a $CDR$ of 1 (0 dB). In reality, hydrometeors are not perfect spheres and chaff strands are not perfect dipoles, but the $CDR$ for chaff is still typically expected to be (and has been observed to be) about 20 dB greater than the $CDR$ for most hydrometeors \cite{moninger+87}.

Radars that operate with SHV-polarized signals cannot directly measure $CDR$. Although \cite{jameson87} showed that $CDR$ can be expressed in terms of linearly polarized radar parameters, \cite{ryzhkov++17} noted that this formulation is quite sensitive to propagation effects, and instead proposed the \pcdr{} approximation, or what they called depolarization ratio ($DR$), that is more robust against propagation effects. \pcdr{}, which is expressed via (\ref{eq.pcdr}), is computed from the measured SHV parameters that can be written as

\begin{equation}
Z_{DR}=\frac{\left\langle\left|s_{\mathrm{hh}}\right|^2\right\rangle}{\left\langle\left|s_{\mathrm{vv}}\right|^2\right\rangle}
\label{eq.zdr_s}
\end{equation}

\noindent{and}

\begin{equation}
\rho_{HV}=\frac{\left\langle s_{\mathrm{hh}}^* s_{\mathrm{vv}}\right\rangle}{\sqrt{\left\langle\left|s_{\mathrm{hh}}\right|^2\right\rangle\left\langle\left|s_{\mathrm{vv}}\right|^2\right\rangle}} ,
\label{eq.rhohv_s}
\end{equation}

\noindent{where $s_{hh}$ and $s_{vv}$ are scattering coefficients for the horizontal and vertical polarizations, respectively, and the asterisk denotes complex conjugation.}

Would \pcdr{} be as good a discriminator between hydrometeors and chaff as $CDR$? For hydrometeors, \cite{ryzhkov++17} showed \pcdr{} to be generally less than -15 dB. For chaff, one would expect \pcdr{} to be close to the theoretical 0 dB for $CDR$, but it is not immediately obvious from (\ref{eq.pcdr}) that this would be the case. \zdr{} and \rhohv{} depend on the flutter angle of chaff \cite{zrnic+04}; although the Hertzian dipole model is only strictly valid for chaff lengths much shorter than the radar wavelength, \cite{zrnic+04} showed that its flutter angle dependency is very similar to a thin wire model. The simple dipole model is convenient, because a closed-form solution is available under reasonable assumptions. \blue{A derivation of this model is provided in the Appendix.}

This is visually illustrated in Fig. \ref{fig_flutter}, where the resulting dependency of \pcdr{} on the flutter angle is shown. It is clear that although \pcdr{} has a flutter angle dependency, the range is quite limited (-3.7 to 0 dB). Since hydrometeor \pcdr{} is expected to be generally less than -15 dB, this derived SHV parameter seems promising for separating chaff from hydrometeor signals.  The \pcdr{} model result is substantially different than the $LDR$ model in \cite{zrnic+04} in its limited range, but the histogram in Fig. \ref{fig_zdr_dr} shows values much lower (as low as -18 dB) than the minimum values shown in Fig. \ref{fig_flutter} (approximately -3.7 dB).  We must consider the fact that chaff is not a perfect dipole, is distributed in space, likely consists of inconsistent cut lengths, and can clump \cite{kurdzo++18}, meaning that we can expect lower values in real-life scenarios.  Additionally, the model presented in (\ref{eq.ai}) assumes a maximum flutter angle of 90\degs{}, meaning that the orientation can only reach an equal distribution between horizontal and vertical (i.e., \zdr{} of 0 dB).  If the chaff orientation distribution leans toward the vertical (i.e., negative \zdr{}), the \pcdr{} values will extend lower than what is modeled in Fig. \ref{fig_flutter}, likely partially contributing to the extended tail seen in Fig. \ref{fig_zdr_dr}.

In considering the utility of the distribution shown in Fig. \ref{fig_zdr_dr}, the measured distribution is still considerably narrower than that of hydrometeors, and the majority of the distribution is well above the expected range of hydrometeor \pcdr{} values \cite{kilambi++18}.  This analysis suggests that an estimate of \pcdr{} can be useful for discrimination of chaff using advanced methods such as ML.  While an ML algorithm could conceivably discern the different weights of importance of \zdr{} and \rhohv{}, especially in combination, between chaff and weather, feeding more-primitive ML algorithms a direct relation such as that provided in \pcdr{} may lead to a better/faster convergence on an acceptable classifier.  Indeed, this has been seen in the recent ongoing development of a chaff detector, first presented in \cite{kurdzo++17b}.  With the use of support vector machines, overall performance has been determined to increase in their approach by using a \pcdr{} estimate of WSR-88D data \cite{kurdzo++23}.

\begin{figure}[!t]
\centering
\includegraphics[width=3.5in]{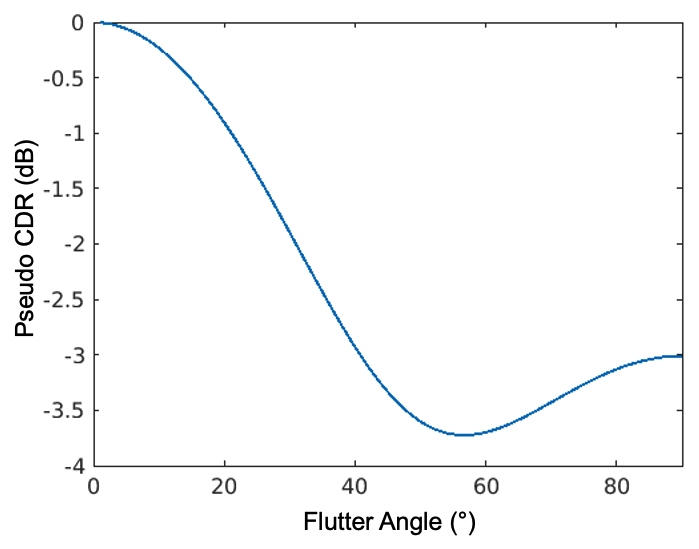}
\caption{$PCDR$s for a dipole model based on flutter angle.  Compared with the $LDR$ results in \cite{zrnic+04}, it is clear that the expected values of \pcdr{} are higher than $LDR$ for a perfect dipole, even at extreme flutter angles.  In fact, the entire range of \pcdr{} is only about 3.7 dB for the a dipole covering all flutter angles.}
\label{fig_flutter}
\end{figure}

\subsection{Height Dependence of \zh{}, \zv{}, \zdr{}, and \pcdr{}}

Given the importance of driving a ML-based algorithm with relevant data that can lead to convergence on a better model more quickly, it is relevant to explore the spatial statistics of chaff.  This is particularly relevant because spatial variables (range, height, etc.) are readily available in real-time processing, and any potential signal that can be derived using a large dataset could be simply included in a ML algorithm (both in training and operations).  In order to identify these potential signals in the present dataset, four variables are examined by height: \zh{}, \zv{}, \zdr{}, and \pcdr{}.  Relationships with range were also explored, but are not shown in this paper.

There is little reason to expect a \textit{direct} dependence of chaff characteristics as a function of range alone (other than a decrease in sensitivity), but there are three other spatially oriented aspects that could theoretically have impacts on what the radar estimates.  These three aspects are height, resolution volume size (which is correlated with range), and elevation angle of the antenna.  Height dependence could arguably be a physically viable expectation due primarily to the results in \cite{kurdzo++18}.  For example, different stages of chaff dispersal, as well as variations in the effect of the electric field (e.g., by height and/or by chaff cut length) could have impacts on statistical distributions.  According to \cite{kurdzo++18}, chaff is often more likely to be vertically oriented at lower elevations due to the increasing strength of the fair weather electric field.

Resolution volume dependence could be viable because as the volume increases in size, it is less likely that chaff will meet the volume-filled scattering assumption in the calculation of reflectivity factor \cite{doviak+93}.  Finally, the elevation angle of the antenna could have an impact due to the fact that the ``vertical'' polarization of the radar becomes less truly vertical as the elevation angle increases.  This is because as the antenna points at higher elevations, the angle of the vertical polarization becomes less orthogonal relative to the ground (becoming parallel to the ground when pointing to 90\degs{}).

In terms of viability, it seems unlikely that the impact of elevation angle would be significant with the WSR-88D due to the relatively low antenna pointing angles (up to 19.5\degs{} maximum, but focused mostly at lower angles, especially in clear air VCPs).  It also seems unlikely that the resolution volume size is a meaningful factor, since we see the opposite effect in Fig. \ref{fig_z_zdr_dr_height}.  That is, reflectivity increases with height rather than decreasing like we would expect if the resolution volumes were not being adequately filled to meet the reflectivity factor calculation assumption.

Therefore, we are left with exploring the impacts of height dependence based on different stages of fallout.  The subplots in Fig. \ref{fig_z_zdr_dr_height} allow for the exploration of these variations, including increasing \zh{} and \zv{} with height, increasing \zdr{} with height, and a slightly decreasing \pcdr{} with height.  When chaff is initially released, it appears in a relatively small footprint on the WSR-88D before dispersing with the prevailing atmospheric wind.  This dispersion results in size sorting vertically \cite{kurdzo++18}, with clumps and vertically oriented fibers falling out prior to horizontally oriented fibers.  Before widespread dispersion spatially, the concentration of chaff is at its highest point immediately post-release.  Therefore, it makes intuitive sense that \zh{} and \zv{} would be at their highest values at higher heights (near release points), and decrease at lower elevations where the originally concentrated chaff has dispersed significantly due to the effects of differential advection (in the horizontal) and differential sedimentation (in the vertical).  Since sedimentation only operates in one direction (downward), it makes sense that the further chaff gets from the height of release (further downward), the weaker the reflectivity factor will be on average.

Additionally, the results presented in \cite{kurdzo++18} suggest that lower \zdr{} values are likely at lower heights (due to the impact of the fair weather electric field), which agrees well with the \zdr{} distribution seen in Fig. \ref{fig_z_zdr_dr_height}.  This is because longer chaff strands are more-heavily impacted by the increasing electric field strength closer to the Earth's surface, causing them to tilt more vertically and hence fall out faster.  An additional consideration not mentioned in \cite{kurdzo++18} that has become evident in the current study is that there may be holistic effects on \zdr{} as chaff descends into the planetary boundary layer (PBL).  Turbulent mixing within the PBL may change the chaff orientation distribution from one that is (for example) mostly horizontal to a more isotropic distribution, which would shift the \zdr{} distribution from positive values toward values closer to 0 dB near the Earth's surface.

\blue{One concern regarding the suggested PBL impacts is that there may be unintended effects within the PBL in the chaff database.  For example, if not carefully considered, chaff in the presence of other target types typically found within the PBL (clutter, insects, and other biota) could affect the statistics.  The truthing mechanism used in this study mitigates this concern as much as possible.  Given that each ``cell'' of chaff was manually traced using a region-of-interest polygon tool, it was determined by a subject matter expert that the gates being included were \textit{dominated} by chaff.  This is an important distinction, since clearly chaff can be mixed with other scatterers.  In practice, if it was not exceedingly clear to the subject matter expert that the dominant scatterer was chaff, the case was not included in the database.  While it is still possible that some biota and other targets are included, it is expected that the influence is exceptionally small.}

Finally, the slight decrease in \pcdr{} is difficult to explain physically.  The model depicted in Fig. \ref{fig_flutter} suggests that as the flutter angle decreases (a more horizontally oriented distribution), \pcdr{} \textit{generally} increases.  This would suggest that we should see increasing \pcdr{} with height since \zdr{} is increasing with height.  It should be considered that \rhohv{} is not being taken into account when simply comparing against flutter angle (a \rhohv{} dipole model is not shown in this paper).  Additionally, as mentioned above, inconsistencies with chaff distribution, cut lengths, and clumps result in \pcdr{} values well outside the range predicted by the simplistic Hertzian dipole model.  With these issues taken into account, it is not clear whether the shift in \pcdr{} with height is statistically meaningful or if there is even an underlying physical mechanism at play.  A more-realistic model for \pcdr{} in chaff would be required to determine the answers to these questions, an endeavor that is beyond the scope of this paper.  What is important in these findings is that the distribution of \pcdr{} with height is remarkably narrow and consistent, making this parameter an excellent candidate for chaff identification.

\red{One concern with the generalization of these statistics is that chaff is not released in a systematic way.  That is, releases do not seem to correlate spatially, are not released in the same densities or at the same heights, and of course, occur in different atmospheric conditions.  As shown in \cite{kurdzo++18}, there is certainly a case dependence, so it must be stated that large-scale generalization has risks.  However, after years of watching chaff releases on the WSR-88D network, the authors have observed patterns in these releases that are generalizable in, at least, qualitative terms.  This is indicated by the similarities in fallout patterns in the 8 cases shown in \cite{kurdzo++18}, which was the impetus for the present study.  Although qualitative generalization is helpful, it is assumed that with millions of gates of chaff and hundreds of cases, quantitative generalization is also both possible and helpful.  In the case of the data presented in Fig. \ref{fig_z_zdr_dr_height}, the discussion in this section presents several theories for why these statistics are physically plausible.}

These are generally unsurprising results given previous findings.  However, the implications cannot be understated.  In a ML algorithm that does not take height (or any spatial considerations) into account, comparing two areas of statistical data at two very different heights could result in the same inference, resulting in an incorrect classification.  The findings in this study helped the authors realize the importance of including spatial dimension(s) in ML approaches to classification, as is described in \cite{kurdzo++23} compared to the original approach discussed in \cite{kurdzo++17b}.  It should also be noted that \textit{generalizations} from a large dataset are necessary for effective ML algorithm training.  While these results generally match those from the case studies in \cite{kurdzo++18}, case studies are not sufficient to train an expansive algorithm.  Discovery of these trends on a larger scale is important for such classification problems.

\subsection{VCP Dependence of \zh{}}

In searching for additional methods for improving the ML techniques used in \cite{kurdzo++17b}, the authors noticed that in several cases, the statistics in chaff clouds \red{seemed to change} qualitatively when the VCP of the radar changed.  This led to the desire to a) quantify these changes, and b) determine the potential applications for algorithm development.  As with the importance of spatial statistics mentioned in the previous subsection, changes in statistics by VCP can cause incorrect classifications if not included in training an algorithm.  While \red{qualitative} statistical changes were observed for several variables, \red{the most quantifiable change with regard to large-scale statistics was seen in the \zh{} histograms, which are} presented in Fig. \ref{fig_z_zdr_vcp}.  The histograms for \zh{} are broken into \blue{two} sets of VCPs: all precipitation VCPs \blue{and all clear air VCPs}.

\red{Two possible explanations were brought to light during this analysis.  First, one of the clear-air VCPs, VCP 31, utilizes a longer pulse, leading to higher sensitivity.  However, this higher sensitivity only lowers the minimum detectable \zh{}, meaning the noise floor is effectively decreased.  If VCP 31 was sufficiently represented in the database, and there were a large number of gates that were below the noise floor of the other VCPs, it would be plausible to think that VCP 31's sensitivity could shift the distributions in this way.  A cursory analysis (not shown) of the spatial distribution of chaff clouds in the database combined with an assumed chaff \zh{} distribution and the performance metrics of each VCP showed that the shift in distribution was negligible between the scan modes.  In fact, the peak did not shift by more than 1 dBZ, leading to the need for another explanation.}

\red{It turns out that the database is heavily weighted by clear-air observations, since chaff is \textit{mostly} observed away from weather.  One of the hallmarks of clear-air VCPs is that they only scan the lower tilts, leaving the higher tilts out in favor of slower scan speeds while maintaining a respectable volumetric update rate.  When so many cases in the database only contain the lower tilts, one must consider the height-based distributions in Fig. \ref{fig_z_zdr_dr_height} that show decreasing \zh{} at lower elevations above radar level.  The difference in precipitation versus clear-air VCPs in Fig. \ref{fig_z_zdr_vcp} are likely a manifestation of the difference in height weighted by the percentage of cases collected in clear-air VCPs.}

\red{It should be noted that there can also be other factors biasing the dataset in this manner.  We do not consider initial chaff release height, how long it stays in fallout/dispersion, the distribution of heights/ranges versus the available tilts, etc.  There are far too many levers to pull that would be potentially better explained by studies directed more at the physical properties of chaff, such as T-matrix calculations.  What is important here is that ML algorithms, in their development, training, and testing/validation, must take into account these differences in one way or another.  In some cases, the data used in training inherently include these biases; however, in the case of chaff and weather radar, this is another example for the need for spatial data modalities in model training and validation.}

\section{Conclusions}
In this study, an analysis of statistics in a new database of chaff cases was completed using the WSR-88D weather radar network.  A total of 267 cases were used, spanning 71 sites across the contiguous United States (CONUS).  This dataset is unique in that it includes an expanded \zdr{} range from \nobreakdash-7.9 -- 7.9 dB to -13 -- 20 dB.  Additionally, several new parameters were investigated that have not yet appeared in the literature, including a proxy for circular depolarization ratio (\pcdr{}), an analysis of \zh{}, \zv{}, \zdr{}, and \pcdr{} by height, and an analysis of \zh{} and \zdr{} statistics by VCP.  The anticipated physical mechanisms for these new observations were discussed, and their relation to algorithm development, especially using ML techniques, was focused upon.

The primary new findings are that \zdr{} values extend all the way to the edges of the new \zdr{} range, but there is a wider spread in the histogram at strongly positive values of \zdr{}, especially between 8 and 13 dB.  \pcdr{} values peak at -4 dB, which is close to the value theorized in an adapted \pcdr{}-based model derived from the $LDR$-based Hertzian dipole model in \cite{zrnic+04}.  However, they trail off toward -20 dB, which indicates expected differences between chaff and true dipoles.  The separation relative to hydrometeors is still sufficient to be an excellent separator between chaff and weather, but it is not expected that it will help as much in differentiating chaff from other statistically similar targets such as sea clutter.  While the results of \zh{} and \zdr{} distributions with height are not necessarily surprising, they demonstrate the need to include spatial variance in statistical distributions in radar algorithms, especially those that use statistics in clusters with ML techniques.  A similar conclusion can be drawn using the VCP findings, which argue that different radar scanning modes should also be included in ML algorithms.

\red{It should be clearly stated that some of the discussion is speculative, without full evidence for the hypotheses stated.  While this may not be fully scientifically satisfying, the hope is that the discussion will spur additional research in these areas, specifically with the thinking that chaff could actually \textit{help} in making new scientific discoveries.  A few examples would be investigating PBL structure and electrification using controlled chaff releases, as well as the potential to release chaff nearby supercell thunderstorms in order to examine air motion in and around severe thunderstorms \cite{moninger+87}.}

\blue{Moving forward, in the future, the findings in this study should be generalized to other radar systems, processing chains, and operating frequencies.  Application of the results in this study to non-WSR-88D radars is recognized as a difficulty, since different radars/frequencies/calibrations could have major effects on the distributions.  Generalization would allow for the ability to detect (and possibly filter) chaff across other radar networks.  There are multiple ways to accomplish this, including manually collecting chaff cases on other networks/systems, or possibly by conducting T-matrix scattering simulations in order to generalize the polarimetric characteristics of chaff at multiple frequencies.  It would also be useful to examine signal-level (I/Q) data within chaff at multiple bands in order to determine any specific spectral characteristics and/or attempt different processing schemes/algorithms.}

\blue{Finally, it would be prudent to investigate cases of chaff mixed with weather as a specific focus in future work.  Although this was not seen often, distributions (and detection) in these cases would be useful to the community.  Building a database of these cases would be challenging and would likely need to take place over multiple years (and with more time steps per case) due to their scarcity.  However, when chaff is located near weather, its contamination is certainly a problem for weather radar users, making this an interesting area for further exploration.}

\red{Specifically, spectral polarimetry is one potential method to separate chaff from weather in a mixed resolution volume.  This approach would require I/Q data to explore, which is something that was not readily available for this study.  Future I/Q collection in chaff could open the door to exploring this issue, although it is still not expected that clean separation can be achieved since chaff and weather occupy the same Doppler spectral bins (i.e., chaff is a passive tracer).}

\appendix[\blue{Derivation of Hertzian Dipole Model for PCDR}]

\blue{Assuming negligible multiple scattering and consistent chaff cut length, the Hertzian dipole model yields \cite{zrnic+04}:}

\begin{subequations}

\begin{equation}
\begin{aligned}
& \left\langle\left|s_{\mathrm{hh}}\right|^2\right\rangle=\left\langle\left|f_b\right|^2\right\rangle-\left\langle\left(f_b^2-f_a f_b\right) A_2\right\rangle+ \\ 
& \left\langle\left|f_b-f_a\right|^2 A_4\right\rangle
\end{aligned}
\end{equation}

\begin{equation}
\begin{aligned}
& \left\langle\left|s_{\mathrm{vv}}\right|^2\right\rangle=\left\langle\left|f_b\right|^2\right\rangle-2\left\langle\left(f_b^2-f_a f_b\right) A_1\right\rangle \\
& +\left\langle\left|f_b-f_a\right|^2 A_3\right\rangle
\end{aligned}
\end{equation}

\begin{equation}
\begin{aligned}
& \left\langle s_{\mathrm{hh}}^* s_{\mathrm{vv}}\right\rangle=\left\langle\left|f_b\right|^2\right\rangle+\left\langle\left|f_b-f_a\right|^2 A_5\right\rangle \\
&-\left\langle\left(f_b-f_a f_b\right) A_1\right\rangle-\left\langle\left(f_b^2-f_a f_b\right) A_2\right\rangle,
\end{aligned}
\end{equation}

\label{eq.hdm}

\end{subequations}

\blue{\noindent{where $f_a$ and $f_b$ are the scattering amplitudes for the electric field along and perpendicular to the dipole axis, respectively. The $A_{i}$ values are the scattering amplitudes assumed to be fixed with a transverse amplitude of $f_b=0$.  These amplitudes are closed-form solutions for the angular moments as described in Eqs. 2--6 in \cite{zrnic+04}.  Furthermore, if the scattering amplitudes are assumed constant and the transverse component is zero ($f_b$ = 0), then (\ref{eq.hdm}) reduces to}}

\begin{subequations}

\begin{equation}
\begin{aligned}
\left\langle\left|s_{\mathrm{hh}}\right|^2\right\rangle=\left|f_a\right|^2\left\langle A_4\right\rangle
\end{aligned}
\end{equation}

\begin{equation}
\begin{aligned}\left\langle\left|s_{\mathrm{vv}}\right|^2\right\rangle=\left|f_a\right|^2\left\langle A_3\right\rangle
\end{aligned}
\end{equation}

\begin{equation}
\begin{aligned}\left\langle s_{\mathrm{hh}}^* s_{\mathrm{vv}}\right\rangle=\left|f_a\right|^2\left\langle A_5\right\rangle.
\end{aligned}
\end{equation}

\label{eq.hdm_red}

\end{subequations}

\blue{\noindent{Substituting (\ref{eq.hdm_red}) into (\ref{eq.zdr_s}) and (\ref{eq.rhohv_s}) yields}}

\begin{equation}
Z_{DR}=\frac{\left\langle A_4\right\rangle}{\left\langle A_3\right\rangle}
\label{eq.zdr_a}
\end{equation}

\blue{\noindent{and}}

\begin{equation}
\rho_{HV}=\frac{\left\langle A_5\right\rangle}{\sqrt{\left\langle A_3\right\rangle\left\langle A_4\right\rangle}}.
\label{eq.rhohv_a}
\end{equation}

\blue{\noindent{Substituting (\ref{eq.zdr_a}) and (\ref{eq.rhohv_a}) into (\ref{eq.pcdr}) gives}}

\begin{equation}
PCDR=\frac{A_3+A_4-2 A_5}{A_3+A_4+2 A_5}.
\label{eq.pcdr_a}
\end{equation}

\blue{Assuming that the chaff is randomly oriented in the horizontal plane, the radar antenna elevation angle is 0\degs, and the angle between the chaff axis and horizontal plane is uniformly distributed between 0\degs{} and a maximum angle dubbed the flutter angle, \cite{zrnic+04} derive closed form solutions for angular moments $A_i$ as a function of the flutter angle's complement $\theta_1$ (i.e., the flutter angle is $\pi$/2 - $\theta_1$, and $\theta_1$ is measured with respect to the true vertical):}

\begin{subequations}

\begin{equation}
\begin{aligned}
\left\langle A_3\right\rangle=\frac{1}{5} \cos ^4 \theta_1
\end{aligned}
\end{equation}

\begin{equation}
\begin{aligned}
\left\langle A_4\right\rangle=\frac{3}{40}\left(\sin ^4 \theta_1-\frac{4}{3} \cos ^2 \theta_1+4\right)
\end{aligned}
\end{equation}

\begin{equation}
\begin{aligned}
\left\langle A_5\right\rangle=\frac{1}{2}\left(\frac{1}{3} \cos ^2 \theta_1-\frac{1}{5} \cos ^4 \theta_1\right).
\end{aligned}
\end{equation}

\label{eq.ai}

\end{subequations}


%



\section*{Acknowledgment}
The authors thank three anonymous reviewers for their helpful comments in revising the original manuscript.  This material is based upon work supported by the Observations Program within the NOAA/OAR Weather Program Office under Award No. NA21OAR4590385.  Any opinions, findings, conclusions, or recommendations expressed in this material are those of the authors and do not necessarily reflect the views of the National Oceanic and Atmospheric Administration.  The authors would like to thank Autumn Millard for her assistance with this work during her internship in the Summer of 2022.  Discussions with Earle Williams, David Smalley, Mark Veillette, David Patterson, Dusan Zrnic, Valery Melnikov, and Alexander Ryhzkov have been exceedingly helpful over the past several years as we continue our desire to learn more about how chaff interacts with weather radar systems.

\ifCLASSOPTIONcaptionsoff
  \newpage
\fi



%


\bibliographystyle{IEEEtran}
\bibliography{Kurdzo_Bibliography}

%

\begin{IEEEbiography}[{\includegraphics[width=1in,height=1.25in,clip,keepaspectratio]{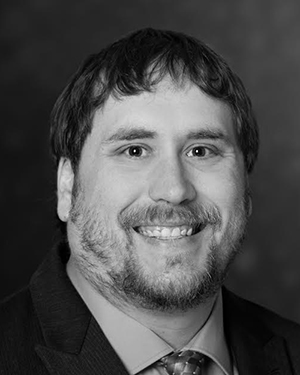}}]{James M. Kurdzo}
James M. Kurdzo was born in Derby, CT, USA, in 1987.  He received the B.S. degree in meteorology from Millersville University, Millersville, PA, USA in 2009, as well as the M.S. degree in meteorology, the M.S. degree in electrical and computer engineering, and the Ph.D. degree in meteorology from the University of Oklahoma, Norman, Oklahoma, USA, in 2011, 2013, and 2015, respectively.

He has been a Technical Staff member at MIT Lincoln Laboratory since 2016, where he is part of the Air Traffic Control Systems Group.  He has authored over 30 refereed publications and over 100 conference papers/presentations covering a range of research interests, including radar network and waveform design, severe convective thunderstorms and tornadoes, phased array weather radar technologies, and weather radar algorithms.

Dr. Kurdzo is a senior member of the Institute of Electrical and Electronics Engineers (IEEE) and a member of the American Meteorological Society (AMS). He was a recipient of the MIT Lincoln Laboratory Early Career Technical Achievement award in 2022, the Tommy C. Craighead award for Best Paper in Radar Meteorology (Univ. of Oklahoma) in 2015, and several presentation/paper awards at AMS conferences.
\end{IEEEbiography}


\begin{IEEEbiography}[{\includegraphics[width=1in,height=1.25in,clip,keepaspectratio]{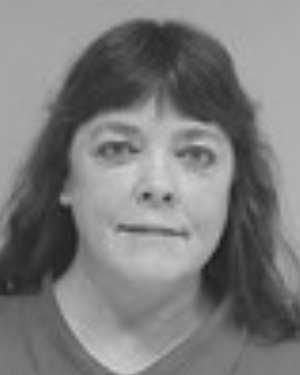}}]{Betty J. Bennett}
Betty J. Bennett was born in Milwaukee, WI, USA, in 1963.  She received the B.S. degree in computer science from Northern Michigan University, Marquette, MI, USA, in 1986.  Since joining MIT Lincoln Laboratory in 1988, Betty has worked in a variety of areas with a focus on real-time systems.  She is currently a member of the Air Traffic Control Systems Group at Lincoln Laboratory working on the NEXRAD Product Enhancement Team.  She has authored or co-authored over 30 refereed articles, technical reports, and conference papers/presentations.
\end{IEEEbiography}

\vfill


\begin{IEEEbiography}[{\includegraphics[width=1in,height=1.25in,clip,keepaspectratio]{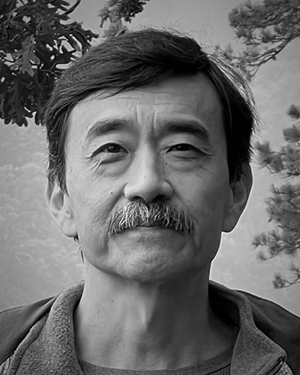}}]{John Y. N. Cho}
John Y. N. Cho was born in Tokyo, Japan, in 1963. He received the B.S. and M.S. degrees in electrical engineering from Stanford University, Stanford, CA, USA, in 1985 and 1986, respectively, and the Ph.D. degree in electrical engineering from Cornell University, Ithaca, NY, USA, in 1993.

From 1986 to 1988, he was a U.S. Peace Corps Volunteer in Sierra Leone. From 1993 to 1997, he was a Research Associate at the National Astronomy and Ionosphere Center's Arecibo Observatory, Arecibo, PR, USA. In 1996, he joined the Leibniz Institute for Atmospheric Physics, K$\ddot{\textup{u}}$hlungsborn, Germany, as a Visiting Scientist. From 1997 to 2002, he was a Research Scientist with the Department of Earth, Atmospheric, and Planetary Sciences, Massachusetts Institute of Technology, Cambridge, MA, USA. Since 2002, he has been with the MIT Lincoln Laboratory, Lexington, MA, USA, where he is currently a Senior Staff with the Air Traffic Control Systems Group. He has authored or co-authored over 60 refereed articles and over 60 conference papers and technical reports. His research interests include atmospheric radar algorithms, weather sensor benefit quantification, weather effects on air traffic flow management, atmospheric waves and turbulence, noctilucent clouds, and meteors.

Dr. Cho is a member of the American Geophysical Union and the American Meteorological Society. He was a recipient of the CEDAR Prize (NSF), the Young Scientist Award (URSI), and the Best Paper Award (10th USA/Europe ATM R\&D Seminar).
\end{IEEEbiography}

\begin{IEEEbiography}[{\includegraphics[width=1in,height=1.25in,clip,keepaspectratio]{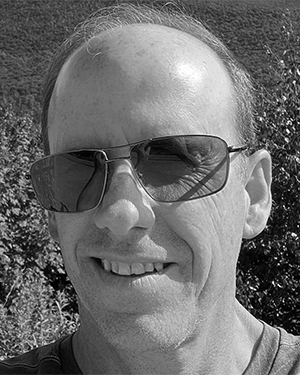}}]{Michael F. Donovan}
Michael F. Donovan was born in Hudson, MA, USA, in 1963. He received the B.S. in meteorology and atmospheric science from University of Massachusetts, Lowell, MA, USA, in 1985. He is currently an Assistant Staff member at MIT Lincoln Laboratory and has been a member of the Air Traffic Control Systems Group since 1987. He has authored or co-authored over 10 refereed articles and over 20 conference papers and technical reports. His research interests include weather radar algorithms, as well as aviation hazards associated with convective thunderstorms and in-flight icing. He is a member of the American Meteorological Society.
\end{IEEEbiography}


\enlargethispage{-5in}

\end{document}